\documentclass[]{vgtc}                 

\ifpdf
  \pdfoutput=1\relax                   
  \usepackage{graphicx}                
  \DeclareGraphicsExtensions{.pdf,.png,.jpg,.jpeg} 
\else
  \usepackage{graphicx}                
  \DeclareGraphicsExtensions{.eps}     
\fi%

\graphicspath{{figures/}{pictures/}{images/}{./}} 

\usepackage{microtype}                 
\PassOptionsToPackage{warn}{textcomp}  
\usepackage{textcomp}                  
\usepackage{mathptmx}                  
\usepackage{times}                     
\usepackage{cite}                      
\usepackage{tabu}                      
\usepackage{booktabs}                  

\onlineid{9740}

\vgtccategory{Subject Development}

\vgtcinsertpkg

\title{What We See and What We Get from Visualization:\\ Eye Tracking Beyond Gaze Distributions and Scanpaths}

\author{Kuno Kurzhals\thanks{email:kunok@ethz.ch}\\
\scriptsize ETH Zurich
\and Michael Burch\thanks{email:m.burch@tue.nl}\\
\scriptsize Eindhoven University of Technology
\and Daniel Weiskopf\thanks{email:daniel.weiskopf@visus.uni-stuttgart.de}\\
\scriptsize University of Stuttgart
}

\teaser{
\centering
  \includegraphics[width=11.5cm]{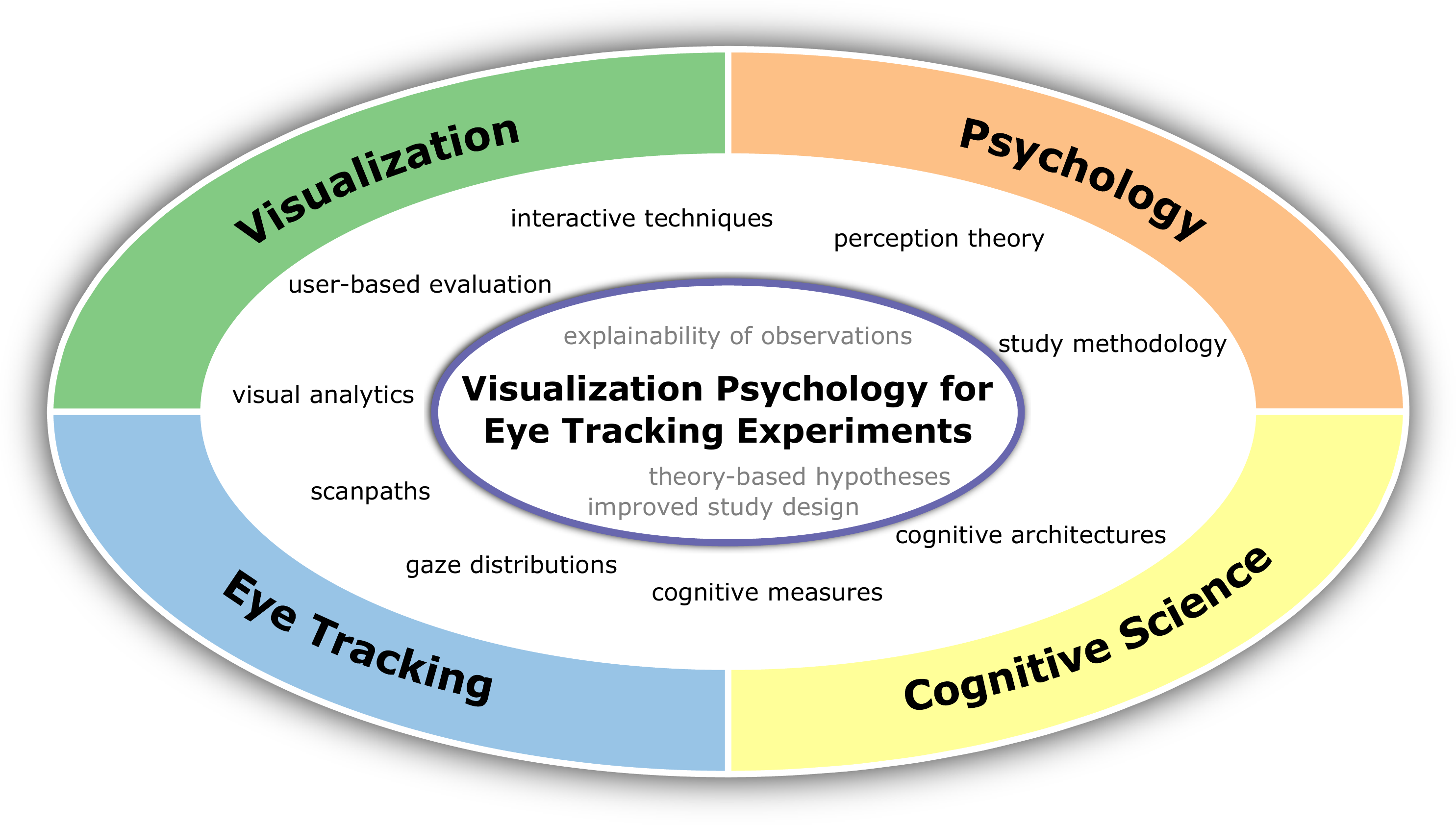}
  \caption{Visualization psychology for eye tracking experiments incorporates expertise from psychology and cognitive science to improve the evaluation of visualization techniques by study methodology, theory integration, and cognitive architectures. 
  }
  \label{fig:teaser}
}

\abstract{Technical progress in hardware and software enables us to record gaze data in everyday situations and over long time spans. Among a multitude of research opportunities, this technology enables visualization researchers to catch a glimpse behind performance measures and into the perceptual and cognitive processes of people using visualization techniques. The majority of eye tracking studies performed for visualization research is limited to the analysis of gaze distributions and aggregated statistics, thus only covering a small portion of insights that can be derived from gaze data. 
We argue that incorporating theories and methodology from psychology and cognitive science will benefit the design and evaluation of eye tracking experiments for visualization.
This position paper outlines our experiences with eye tracking in visualization and states the benefits that an interdisciplinary research field on visualization psychology might bring for better understanding how people interpret visualizations.
} 

\CCScatlist{
  \CCScatTwelve{Human-centered computing}{Visu\-al\-iza\-tion}{Visualization design and evaluation methods}{}
}

\begin{document}


\firstsection{Introduction}

\maketitle

Eye tracking experiments in visualization research have provided insights into how people interpret and interact with visualizations. 
In contrast to classic performance analysis, the analysis of gaze behavior provides information about the distribution of visual attention over time and visual strategies employed in  interpreting a visualization or in working with a complex visual analytics system. 
Typical measures derived from gaze data are fixation duration, fixation count, saccade length, and numerous other aggregated values~\cite{Holmqvist2011}. All measures can be indicators for specific perceptual or cognitive aspects (e.g., cognitive load~\cite{vanGog2009}, working memory~\cite{Padilla2020}) that are potentially interesting for the assessment of a visualization. 

Previously, we surveyed the types of eye tracking studies performed in visualization research~\cite{kurzhals2016} and noticed that mainly gaze distributions, fixation sequences, and comparisons between user groups were analyzed. Back then, we argued that eye tracking can serve \textit{``as a method and data source that can
be interpreted from both psychological and visualization perspectives, acting as a bridge between cognitive and computing science''} with the long-term goal to build an interdisciplinary scientific community. Since then,  collaborations between communities have increased, for example, reflected in several activities at IEEE VIS\footnote{https://visxvision.com}$^,$\footnote{https://decisive-workshop.dbvis.de}$^,$\footnote{https://www.etvis.org}.

Another example is the collaborative research center SFB-TRR~161\footnote{https://www.sfbtrr161.de/} 
on quantitative methods for visual computing, which brings together researchers from both fields in a common interdisciplinary research project.
Such projects are important steps to make progress in the evaluation of visualization methods beyond usability testing.
In addition, machine learning, statistics, visualization research, and data science in general contributed a multitude of new techniques~\cite{blascheck2017,Duchowski2017} to expand the spatio-temporal analysis of eye tracking data, verify results, and formulate new hypotheses.
By combining such state-of-the-art analysis techniques with expertise from psychology, cognitive science, and eye tracking research, as depicted in Figure~\ref{fig:teaser}, the design and insights gained from eye tracking experiments in visualization can be significantly improved.


\section{Experiences}

We summarize our experience from the aforementioned activities in which we were involved and from our observations of the community. 
Hypothesis building in eye tracking studies for visualization is often data-driven, based on observations from pilot studies or previous experiments. Similarly, reported results of eye tracking studies are typically summaries of observations. In both cases, theories of perceptual and cognitive psychology are less prominent, although they might help explain specific observations. We see three main reasons for this situation concerning the research background and interests, and the complexity of visualization problems at hand.

\paragraph{Research Background}

Although many design principles are based on perceptual and cognitive theories, in-depth psychological background knowledge is often not part of the education for visualization. 
Researchers starting with eye tracking studies are confronted with learning eye tracking methodology, which is, starting with proper calibration to a comprehensive analysis of the data, a complex field on its own. As a consequence, deeper knowledge of a whole new research field, i.e., psychology, is hard to achieve within the short time span of an average PhD student's career.

\paragraph{Research Interests}

Psychologists' core topics are often disconnected from topics relevant for visualization research. Yet, there are some successful examples of combining communities,  for example, at the \textit{Symposium on Eye Tracking Research and Applications (ETRA)}. Such events provide great opportunities for interdisciplinary discourse and establishing collaborations. However, publication strategies and research topics might significantly differ between communities.
Hence, a fusion of expertise just by project collaborations might cover some research questions, but from a long-term perspective, other solutions are necessary.



\paragraph{Complexity}

The increasing complexity of visual analytics approaches exacerbates an interpretation of how different perceptual and cognitive processes work together when participants interact with a system. 
Furthermore, study tasks in visualization are often a performance measure to assess the quality of a technique, while in psychological experiments, the task is often just a vehicle to induce cognitive processes.
Consequently, more directed research on visualization-specific aspects is necessary, which conflicts with research on generally applicable theories.


\section{Visualization Psychology for Eye Tracking}

For the aforementioned reasons, we see the need of \textit{visualization psychology} as a new field in visualization research.
Close collaborations with psychologists and cognitive scientists in the education program for visualization researchers promises multiple improvements: 

\paragraph{Improving Study Quality}

Based on cognitive and psychological models, an eye tracking study might be designed in a way that it focuses more on the research questions and the hypotheses in mind. Given the tasks to be solved and the participant group with lots of involved participant-related properties can make it easier to design such a study since it restricts the number of independent variables to be tested. Hence, knowing more about the psychological aspects can improve the study quality and consequently, the reliability and generalizability of the results.


\paragraph{Theory-based Study Design}

A stronger integration of theory in eye tracking study designs has the advantage that the generalizability of the results can be better argued. If one or multiple theories can be linked to specific behavior from interpreting a visualization and interacting with it, the significance of the results is strengthened. Hypotheses derived from observations and supported by theory need to be emphasized in visualization research and should be an essential part of visualization psychology.

\paragraph{Explainability of Observations} 

Vice-versa, study results can be related to theories to explain the observations. For many observations, we can explain \textit{what} happens and \textit{how} something happens. However, another important question is \textit{why} it happens. 
Because many user studies already focus on the first two aspects, visualization psychology for eye tracking should focus on the aspects that can potentially answer why people interpret or interact with a visualization in a specific way.

\paragraph{Cognitive Architectures} 

For many visualization problems, a general understanding of how people process the visual stimulus would be desirable. Consequently, cognitive architectures (e.g., \mbox{ACT-R}~\cite{anderson1997}) that simulate user behavior, based on theoretical concepts and data-driven methods, would help to evaluate new visualization methods, reducing the effort and cost to perform user studies. 

\section{Example Scenario: Metro Maps}
\label{Examples:Sec}

As an example, we discuss one of our eye tracking studies and how it might benefit from visualization psychology.
In a recent eye tracking study, Netzel et al.~\cite{Netzel_SCC:17} describe the results of comparing color-coded and gray-scale public transport maps with the major outcome that color is an important ingredient 
to reduce the cognitive burden to follow lines. The analysis 
showed that in the colored map version the participants had much larger saccades and we hypothesize that the colored lines made them feel safer and hence, the route finding tasks could be answered faster and more reliable. On the other hand, in gray-scale maps, the participants' eyes moved with significantly smaller saccades to trace a line reliably.

The analysis of the eye tracking data focused on gaze distributions, in particular, which regions were investigated how long and in which order.
Due to the lack of a cognitive model that explains the interpretation of such maps, we are not able to argue what someone is thinking at the moment of interpreting the map. 
Finally, we did not include any cognitive models in the eye tracking study to simulate user behavior. This is difficult because the study was one of the first in this domain and  a domain expert served as a means to simulate a person who has a high experience level. However, we did not have a variety of different experience levels and the theoretical background to model participants' behavior in a generalizable way.

\section{Future Directions}
%
Visualization psychology could bridge gaps and strengthen and explain observations from eye tracking experiments in visualization.
It could be a relevant contribution to visualization and psychology research in general, but eye tracking could benefit in particular due to its complexity and the relationship to understanding 
 perceptual and cognitive processes that are involved in interpreting visualizations and working with interactive systems.
 
 A key question, of course, is: \textit{How can we integrate the expertise from both research fields in a common research endeavor?} We think that activities such as this workshop or our own experience with the ETVIS workshop and joint research centers (like SFB-TRR~161) are a good way to go, but are alone not sufficient and need further action. Building a research area of visualization psychology could be a viable means, for example, by establishing publication and other presentation opportunities that work for visualization researchers and psychologists alike, by setting up a canon of teaching new students, and by lobbying for funding possibilities for such interdisciplinary work. 
 We also see the need that visualization researchers should reach out into psychology; so far, much effort has been put into bringing psychology into visualization research, but the long-term strategy should strive for a greater balance between the fields.



\acknowledgments{
This work was funded by Deutsche Forschungsgemeinschaft (DFG, German Research Foundation) -- Project-ID 251654672 -- TRR 161 (Project B01).}

\bibliographystyle{abbrv-doi}

\bibliography{template}
\end{document}